\documentclass[12pt,preprint]{aastex}
\usepackage{amsmath}

\slugcomment{The Astrophysical Journal:  Accepted 11 Nov.  2010 }
\shorttitle{OMC1 outflows}
\shortauthors{Bally et al.}

\begin{document}

\title{Explosive Outflows Powered by the Decay of
Non-Hierarchical Multiple Systems of Massive Stars: \\ Orion BN/KL} 

\author{John Bally}   
\affil{Center for Astrophysics and Space Astronomy, \\ 
       University of Colorado, Boulder, CO 80309}
\email{John.Bally@colorado.edu}

\author{Nathaniel J. Cunningham}
\affil{Department of Physics and Astronomy, University of Nebraska-Lincoln}
\affil{116 Brace Laboratory, Lincoln, NE 68588-0111}
\email{ncunningham2@unl.edu}

\author{Nickolas Moeckel}
\affil{Institute of Astronomy, University of Cambridge}
\affil{Madingley Road, Cambridge CB3 0HA, England}
\email{moeckel@ast.cam.ac.uk}

\author{Michael G. Burton}
\affil{School of Physics, University of New South Wales \\
  Sydney, Australia, NSW 2052}
\email{mgb@phys.unsw.edu.au}

\author{Nathan Smith}
\affil{Astronomy Department, Steward Observatory, University of Arizona, Tucson AZ}
\email{nathans@as.arizona.edu}

\author{Adam Frank}
\affil{Dept. of Physics and Astronomy, \\
       University of Rochester,  Rochester, NY, 14627} 
\email{afrank@pas.rochester.edu}
 
\author{Ake Nordlund}
\affil{Astronomical Observatory and Theoretical Astrophysics Center, \\
       Juliane Maries Vej 30, DK-2100 Copenhagen, Denmark}
\email{aake@astro.ku.dk}


\begin{abstract}
The explosive BN/KL outflow emerging from  OMC1 behind  the  Orion 
Nebula may have been powered by the dynamical decay of a 
non-hierarchical multiple system $\sim$500 years ago that ejected  the 
massive stars I, BN, and source n, with velocities of  about 10 to 30  km~s$^{-1}$.     
New proper motion  measurements  of  H$_2$  features show that within the 
errors of measurement,   the outflow originated from  the site of stellar ejection.  
Combined with published  data, these measurements  indicate an outflow 
age of $\sim$500 years,  similar to the time since stellar ejection.   The total
kinetic  energy  of the  ejected stars and the outflow is about 
$2$ to $6 \times 10^{47}$  ergs.      It is proposed that the gravitational  potential  
energy released by the formation  of a short-period binary, most  likely source I,   
resulted in stellar  ejection and powered the outflow.    
A  scenario is presented for the  formation of a compact,  non-hierarchical 
multiple star system, its decay into  an ejected binary and two high-velocity 
stars,  and launch of the outflow.   Three  mechanisms  may have  contributed to  
the  explosion in the gas:   (i) Unbinding of the  circum-cluster envelope following 
stellar ejection,   (ii) disruption  of  circumstellar disks and high-speed expulsion of 
the resulting debris during the final stellar encounter, and (iii)  the release of stored 
magnetic  energy.   Plausible proto-stellar disk end envelope properties 
can produce the observed outflow mass, velocity,  and  kinetic energy distributions.  
The ejected  stars may have acquired new disks by fall-back or 
Bondi-Hoyle accretion with axes roughly  orthogonal to their velocities.   The 
expulsion of gas and stars  from OMC1 may have  been driven by stellar interactions.  
\end{abstract}

\keywords{star formation: general --- nebulae: individual 
(Orion Nebula, OMC1)}

\section{Introduction}

Massive stars are relatively rare and tend to form in distant, 
obscured, crowded and dynamic environments. The Orion Molecular 
Cloud 1 (OMC1) located immediately behind the Orion Nebula is the 
nearest site of ongoing massive star birth with a bolometric  luminosity 
of about  $10^5$ L$_{\odot}$ (Gezari, Beckman, \& Werner 1998). 
OMC1 contains the Becklin-Neugebauer 
(BN) object (Becklin \& Neugebauer 1967; Scoville et al. 1983), 
the first massive young stellar object to be discovered in the infrared 
and the  Kleinman-Low (KL) thermal-infrared nebula (Kleinmann \& Low 1967)
which contains additional highly obscured massive stars.     In addition to BN, 
radio sources I and n are thought to be massive because they are detected
at radio wavelengths (Menten \& Reid 1995).  Recent sub-millimeter 
interferometry has revealed the presence of an additional 
protostellar candidate, SMA1 about 2\arcsec\ south-south west of 
radio source I  (Beuther et al. 2006; Beuther \& Nissen 2008).    Although
this is the brightest source in the sub-mm, it has not been detected at
other  wavelengths.

The BN/KL region contains a spectacular, wide opening-angle, arcminute-scale  
outflow (Allen \& Burton 1993) traced by molecules such as CO and  NH$_3$ that 
exhibit broad  ($>$ 100 km s$^{-1}$) emission line wings (Kwan \& Scoville 1976; 
Wiseman \& Ho 1996; Furuya \& Shinnaga 2009)  and high-velocity OH, H$_2$O, 
and SiO maser  emission (Genzel et al. 1981;  Greenhill et al. 1998).    This bipolar
outflow with a southeast (red-shifted) - northwest (blue-shifted) axis
contains  at least 8 M$_{\odot}$ of  accelerated gas with a median velocity of about 
20 km s$^{-1}$.     Interferometric CO images, H$_2$O masers,  and dense-gas
tracers such as  SiO reveal smaller (8\arcsec\  long) and younger ($\sim$ 200 
year old) outflow along a  northeast-southwest axis emerging from radio source I  
orthogonal to the arc-minute-scale  CO outflow  (Beuther \& Nissen 2008; Plambeck et al. 
2009).    The  momentum and kinetic   energy content of these flows is at least  
$160$ M$_{\odot}$ km s$^{-1}$ and  $4 \times 10^{46}$  ergs  (Snell et al. 1983) to  
$4 \times 10^{47}$ ergs  (Kwan \& Scoville 1976).     Zapata et al. (2009)  presented a
CO J = 2--1 interferometric study and found a dynamic age of  about $500$ 
years for the larger OMC1 outflow.   They noted its impulsive nature, that its  structure is 
different from accretion-disk  powered  flows,  and that it  originated several arc-seconds 
north of the  OMC1 hot-core.   

While most protostellar outflows  appear to be driven by jets  or collimated 
winds (Reipurth \& Bally 2001; Beuther \& Shepherd 2005;  
Arce et al.  2007; Qiu et al. 2008), the BN/KL outflow is a wide-angle 
explosion.  Near-infrared images reveal hundreds of individual bow shocks  
in the lines of [FeII] and H$_2$ that emerge in nearly every direction  
(Allen \& Burton 1993; Kaifu 2000).     The higher excitation species such 
as the 1.64 $\mu$m line of [FeII] trace the bow shock tips (dubbed ``bullets") 
while lower excitation tracers such as H$_2$  trace the wakes or bow-shock 
skirts (dubbed ``fingers").   A few of these shocks  protrude from the molecular 
cloud and can be seen at visual wavelengths in  [OI], [SII], [NII], 
and H$\alpha$ (Axon \& Taylor 1984; O'Dell et al. 1997).    The highest velocity 
shock, HH~210 ($\sim$ $400$ km~s$^{-1}$) has been  detected in X-rays 
(Grosso et al. 2005).    Proper motions indicate a common age of about $10^3$ 
years or  less for many features (Doi, O'Dell, \& Hartigan 2002), comparable 
to the 500 year age estimate of Zapata et al. (2009).

The three brightest radio-emitting stars in OMC1, sources BN, I, and n, have 
proper motions  (motions in the plane of the sky) of 25, 15, and 26 km s$^{-1}$ 
away from a region less than 500 AU in diameter from which they were ejected 
about  500 years ago (Rodriguez et al. 2005; Gomez et al. 2005; 2008).  
Apparently,  a non-hierarchical multiple star system containing at least 4 
members experienced a dynamical interaction resulting either in the 
formation of a tight binary or possibly even a stellar merger (Bally \& Zinnecker
2005; Bally 2008; Zapata et al 2009) whose (negative) gravitational binding 
energy ejected these stars  from the OMC1 core.  With estimated stellar masses 
of 8 to 13, 12 to 20, and  2 to 3 M$_{\odot}$ for BN, I, and n respectively, the 
kinetic energy of the  ejected stars is $1$ to $2 \times 10^{47}$ ergs 
(Rodriguez, et al., 2005; Gomez et al. 2005;  2008). 
Thus, the total kinetic energy of the ejected stars and the outflow is about 
$1.5$ to $5.5 \times 10^{47}$ ergs.   Assuming that the outflow has suffered
some dissipation of its  original kinetic energy as it interacted with its surroundings,
a reasonable range of initial energies required to eject the stars and outflow
is $2$ to $6 \times 10^{47}$ ergs.    

In this paper, new proper motion measurements of H$_2$ are presented and
combined with published data.  The results indicate a 500 year age for 
the fastest ejecta in the OMC1 bullet and finger system and show that 
the point of origin of the flow coincides with the location of the ejected 
stars  500 years ago to within the measurement errors.  Motivated by an
apparent common event that ejected both the stars and the outflow, we 
present a plausible scenario for the recent evolution of the OMC1 
region.

\section{Observations:  Proper Motions of  Features in the BN/KL Outflow}

New images of the BN/KL outflow were obtained on 21 November 2004 with 
Near Infrared Camera and Fabry-Perot Spectrometer (NICFPS)  on the ARC 3.5 m 
telescope at  Apache Point Observatory in New Mexico.    NICFPS uses a
1024 $\times$ 1024 pixel Rockwell Hawaii 1-RG HgCdTe detector.   The pixel 
scale of this instrument is 0.273\arcsec\ per pixel with a field of view 
4.58\arcmin\ on a side.  Images with 300 second exposures were obtained in 
the 2.122 $\mu$m S(1) line of H$_2$ using a  narrow-band filter 
(FWHM= 0.4\% of the  central wavelength).      Separate sky frames 
were obtained using the same exposure time at a location 600\arcsec\ east.  
A set of 5 dithered images were obtained both on-source and on the sky position.
A median-combined set of unregistered, mode-subtracted sky frames were used
to form a master sky-frame that was subtracted from individual images.    The reduced
images were corrected for optical distortions.   Field stars were used to align the
frames which were  median-combined to produce the image shown in Figure 1.
Atmospheric seeing produced   0.\arcsec 8. full-width at half-maximum stellar images.
A full description of the data acquisition, the data reduction,  and analysis 
procedures is given in Cunningham (2006).

Proper motions for individual features in the BN/KL outflow were determined by 
comparison of the 2004 epoch NICFPS image with H$_2$ images taken on 
13 September 1992  using the IRIS instrument at the Anglo-Australian Telescope 
(AAT)  described by  Allen \& Burton (1993), and the narrowband images  taken on 
11 and 13 January 1999 with the CISCO instrument on the Subaru telescope on 
Mauna Kea, Hawai'i  (Kaifu et al. 2000).   These images were registered to the 
NICFPS image using field stars to an accuracy of about 0\arcsec .05, limited by 
optical distortions.  The  time-intervals between the first image and the second and 
third images are 6.33 and  12.19 years.

Proper motions were determined using difference-squared cross-correlation 
technique of Currie et al. (1996).  All images were re-sampled to the pixel-scale of 
the 2004 epoch NICFPS image and blinked to identify moving features.
Measurement boxes containing between 50 and 500 pixels 
were defined on the 2004 epoch image as shown in Figure~1.
On the prior-epoch images,  a larger measurement box  with 6 additional pixels 
beyond each boundary was defined at each location.   Slowly-varying
background emission was removed by fitting a two-dimensional cubic function 
and subtraction of the result.     The smaller box is shifted within the larger box
for all possible $x$ and $y$ pixel-offsets and subtracted from the larger box.  
The square of the sum of all pixel-to-pixel differences is assigned to the pixel 
that corresponds to the $x$ and $y$ offset in an ``offsets"  image consisting of all possible 
pixel shifts.   The location of the minimum in this two-dimensional array is an 
estimator of the  $x$ and $y$ offset that results in the best match between 
the intensities in the boxes defined on the first and second epoch images.   
Proper motions are estimated using the centroid of a two dimensional gaussian 
fit to this ``offsets" image.   Field stars were used to estimate a mean proper motion 
error of about 15 km~s$^{-1}$ for bright compact sources for both the 2004 -- 1999
and 2004 -- 1992 comparison.

Most proper motion vectors point directly away from the OMC1 cloud core.  
However,  about 10\% of the vectors differ by more than 30\arcdeg\ from the
expected direction of motion away from the point of origin of the outflow.    
The 4 boxes located in the lower-right portion of Figure~1 trace a highly 
collimated chain of H$_2$ bow-shocks that emerge  at position angle 
PA $\sim$ 320\arcdeg\ from  the OMC1-S core located about 90\arcsec\ south OMC1.    
Another collimated  outflow crosses the northwestern lobe of the OMC1 outflow 
at position angle PA $\sim$  240\arcdeg\ to 250\arcdeg .    Its brightest 
southwest-facing bow shock is located at  J(2000) = 
5$^h$35$^m$11.$^s$6, --05\arcdeg\  20\arcmin\  54\arcsec  .
The orientation of this shock indicates that the source is likely to be located
towards the northeast in the ``Integral Shaped Filament"  which contains
the highest column density of molecular gas  (Johnstone \& Bally 1999).

Figure~1 shows  the H$_2$ proper motion vectors.   The fastest
motions are located at the greatest distance from the OMC1 core while the slower
motions are located closer in.  Thus, to first order, the motions are consistent with 
velocity being proportional to distance (e.g. a ``Hubble flow''),  an indication of an 
explosion.   

Figure~2 shows the proper motions of 2.122 $\mu$m H$_2$ features 
measured by Cunningham (2006, and this work), 1.644 $\mu$m [FeII] 
(Lee \& Burton 2000), and Herbig-Haro objects traced by H$\alpha$, 
[OI], [NII], and/or [SII] (Doi, O'Dell, \& Hartigan 2002)  as a function of 
distance from the point of origin of the three high-velocity stars.    
H$_2$ tends to only trace features with relatively low proper motions
below  about 70  km~s$^{-1}$.    Presumably this is consequence of 
H$_2$ dissociation in magnetized C-shocks with speeds in excess of 
about 70 km~s$^{-1}$.     For non-magnetized J-shocks, H$_2$ dissociates 
at shock speeds of 10 -- 30 km~s$^{-1}$.
The pair of $\sim$ 250 km~s$^{-1}$ H$_2$ features located about
0.1 pc from the expansion center in Figure~2 may  trace reverse shocks 
moving into dense molecular clumps  that have high velocities.
H$\alpha$ excitation 
requires shock speeds of at least 60 km~s$^{-1}$ while the $\lambda$ 5007 A
line of [OIII] requires shocks with speed in excess of at least 150 km~s$^{-1}$.
Most visual-wavelength forbidden transitions such as the  
$\lambda$ 6717/6731 [SII] lines trace the cooling layers behind fast shocks.
[FeII] and the visual-wavelength tracers are commonly
associated with features having velocities up to 425 km~s$^{-1}$.
  
Figure~2  demonstrates that some proper-motion features have apparent ages 
between 500 and 1,000 years.    Dynamical ages are determined  
from  the distance traveled divided by the magnitude of the proper motion.   
Unlike age determination from radial velocities, knowledge of the inclination 
of the flow vector is not required.   For example HH~210, the fastest HH object 
associated with the BN/KL outflow, has a dynamic age of 605 years assuming
no deceleration.    Given that most shocks are likely to have been decelerated 
by interactions with their environment, Figure~2 shows that the ages of 
many features in the BN/KL outflow are consistent with ejection about 500 years
ago,   similar to the time elapsed since the ejection of the three high-velocity stars, 
BN, I, and n.   

Figure 3 shows the apparent point of origin of the H$_2$ proper motions in the 
OMC1 outflow.   The left panel is made by backward tracing of the 
proper motion vectors, effectively marking their trajectories in the absence of 
deflections.      The peak in the density of intersecting trajectories provides the
best estimator for the point of origin of the measured motions.     The ``intensity''
of each trail is weighted by the speed.    Thus, large
motions, which can be measured more precisely than small motions,  have a
greater weight.      The image in the right panel has been boxcar-smoothed 
by 31 pixels (25\arcsec ) to aid visualization of the intersection point.    
The point of origin of the proper motions is located at  
J(2000) = 5$^h$~35$^m$~14.$^s$5, --5\arcdeg\  22\arcmin\  23\arcsec , 
with a 1 $\sigma$ uncertainty of about 5\arcsec .   This location is offset by about 5\arcsec\   
to the northeast of the apparent ejection center of the massive stars and 
9\arcsec\ north of the current location of radio source I.    Zapata et al. (2010)
found an expansion center based on interferometric CO J = 2--1  observations
at J(2000) = 5$^h$~35$^m$~14.$^s$37, --5\arcdeg\  22\arcmin\  27.\arcsec 9,   
about 5\arcsec\ due south the expansion center derived from our
proper motions, but consistent within the uncertainty estimates.

\section{Runaway stars, Dynamical Decay,  and the Origin of Explosive Outflows}

High velocity runaway stars are relatively common among  massive  O and 
early B stars  but  rare among lower-mass  stars  (Gies \& Bolton 1986;  Gies 1987).  
While approximately 30\% of O stars are runaways with velocities  greater than 
30 km~s$^{-1}$ with respect to their birth clusters and associations,  less than 
2\% of B and A stars have such velocities (Stone 1991;   Schilbach \& Roser 2008;  
Zinnecker \& Yorke 2007).    Many  (possibly most)  massive runaway stars were 
ejected by  three-body  encounters between a  single star and a massive binary
(Gvaramadze \&  Gualandris  2010),    four-body encounters between a pair of   
massive binaries (Poveda, Ruiz, \& Allen 1967), or the decay of  non-hierarchical 
multiple  star systems (van Albada 1968; Valtonen \& Mikkola 1991; Aarseth et al. 
1994;  Sterzik \& Durisen 1995, 1998).   Such three- or  four-body encounters are most 
likely to occur  in very dense, young star clusters.     The runaway stars AE Aur and
$\mu$ Col,  ejected with speeds of over  100 km~s$^{-1}$ about $2.5 \times 10^6$ years  
ago (Hoogerwerf, de Bruijne,  \& de Zeeuw 2001) from the NGC 1980  cluster 
located about 30\arcmin\  south  of the Orion Nebula,  provide a prototypical  example.   
This event  produced  the massive colliding-wind X-ray binary $\iota$ Ori  
(Gualandris, Portegies-Zwart,  \& Eggleton 2004).     

Non-hierarchical multiple  star systems, triple or  higher-order  
multiples in  which the time-averaged   distances between components  are similar 
are dynamically  unstable.     Numerical  experiments show that non-hierarchical 
systems tend to rearrange  into hierarchical  configurations consisting of a compact 
binary  and either  ejected stars or ones in elongated orbits with a much larger 
semi-major axis  than the binary.   Non-hierarchical systems tend to decay in 
about  100  crossing times;  the  crossing time is roughly given (in years) 
by $t_{cr} \sim 0.17 (R^3 / M)^{1/2}$  where $R$  is the characteristic  
length scale of the system in AU and $M$ is total 
system  mass in Solar masses (Anasova 1986; Sterzik \& Durisen 1995).   
Many low-mass pre-main sequence stars are born in such non-hierarchical 
multiples (Reipurth 2000).    Among massive stars, Trapezium-like groups
tend to be non-hierarchical and subject to dynamical decay.

Both the ejection of massive stars and the  launch of the OMC1 BN/KL outflow  
may be the result of  the dynamic interaction and rearrangement of a system 
of massive stars in OMC1 about 500 years ago.   
The kinetic energy originates from the release  of gravitational 
potential energy accompanying the dynamic formation of a  compact binary.  
According to the Virial theorem, one-half of the (negative) potential energy
of the binary goes into orbital motion of its constituent stars.  The other half
is removed from the system, and is available to accelerate the binary, other
stars involved in the interaction, and surrounding gas.   
The most massive component of the  system,  which is the slowest moving 
object,  is likely to be the  binary.     The binding energy of a binary consisting
of masses $m_1$ and $m_2$ with an orbital semi-major axis $a$,  
$E_b =  - G m_1 m_2 / 2 a$,    must be twice the (negative) sum  of the  
observed kinetic energies in  the outflow and ejected stars.   

Assuming that source I consists of  two 10  M$_{\odot}$ stars,  to  
release  $2 \times 10^{47}$  (or  $6 \times 10^{47}$) ergs  of free kinetic energy,
its members  must  have a semi-major axis  less than 2.2 (or 0.74) AU  with an 
orbital  period shorter than  2  (or 0.4) years. 
The perihelion velocity  of the stars must  be at least  33 (or 55) km s$^{-1}$ and
the binding energy of the binary must be less than $-4   \times 10^{47}$  
(or  $-12 \times 10^{47}$)   ergs. 
If either BN or n is the  binary, the binary separation  must be much smaller 
to provide the energy  for the ejection of the three stars  plus outflow.   
Stellar ejection removed  a total of about 22 to 36 M$_{\odot}$ of stars
and least 8 M$_{\odot}$ of gas. Thus, this event removed  30 to  50 M$_{\odot}$ 
from the OMC1 core.  The point of origin of the ejected stars is located 
at the center of  the BN/KL outflow about 4\arcsec\  ($\sim$ 1700 AU in projection) 
northwest of the  present location  of radio source I.   

Momentum conservation requires that the vector sum of the stellar and outflow 
momenta sum to zero in the rest frame of the parent cloud.  Thus,  the runaway 
single stars and the dynamically formed binary are most likely to be ejected 
from their  star-forming clump.  Removal of the stars implies that the envelope 
whose motion  was dominated by the gravity of the stellar system before ejection 
can become unbound after ejection.  Multi-body encounters tend to eject the outer 
portions  pre-existing circumstellar disks at  radii larger than about 1/3 to 1/2 of the 
periastron separation.   Thus, the outer parts of disks around the stars that form  
the binary and the envelope surrounding the disrupted cluster will be ejected 
with roughly their respective orbit speeds.
  
Three  distinct mechanisms may be involved in releasing gas from the 
gravitational potential well.     First, the disruption and ejection of  the outer-parts 
of  circumstellar  disks during the final dynamic encounter leading the binary formation.
Second, recoil of the pre-existing circum-cluster  envelope following stellar ejection.  
Finally,  the  release of  stored magnetic  stress.    
Each process is a consequence of  the removal of gravitational attraction on 
a time-scale given by the relevant crossing time.    Disk disruption occurs within 
a  time-scale $t_{cross} \approx  r / V << $ 1 year at $r$ = 1 AU for
$V \sim$ 100 km s$^{-1}$, the Kepler speed around a 10 M$_{\odot}$ star.   
At $r$ =  100 AU,  stellar ejection with a mean speed of  $V$ =  20  km~s$^{-1}$ 
implies that the envelope gas would be unbound several decades.   
In both cases, the expulsion  velocity would be comparable 
to the Kepler speed  prior to  ejection, or about 20 km~s$^{-1}$ for an envelope 
initially orbiting 40 M$_{\odot}$ of stars at $r$ = 100 AU to over 100 km~s$^{-1}$ 
for parts of disks ejected from within 1  AU of a 10 M$_{\odot}$ star.
These mechanisms are explored in more detail below.

\subsection{Formation of a Non-Hierarchical Massive Multiple in OMC1}

OMC1 spawned at least 20 to over  40 M$_{\odot}$ 
of stars.   Assuming a typical star-formation efficiency of 25\% to 50\% 
typical of massive cluster  forming clumps,   the OMC1 proto-cluster clump 
must have initially contained at least $M_{clump}$ $\sim$  100  M$_{\odot}$ 
of gas.    As a fiducial example, consider a clump with an `isothermal' density 
distribution $\rho (r) = \rho_0 (r/r_0)^{-2}$  for radius r less than the clump outer
radius $R$.    A  clump with mass $M_{clump} = 100$ M$_{\odot}$ inside 
$R$ = 0.1 pc implies mean orbital velocity (which is independent of $r$),
$V \approx (GM_{clump}/R)^{1/2} = (4 \pi G \rho_0)^{1/2} =   2.1$ km~s$^{-1}$ 
where $\rho_0 \approx 5.0 \times 10^{16}$ g~cm$^{-3}$ would be the
density at $r_0$ = 1 cm if the power-law distribution were extended to this
radius.   In this example,  the density at $r$ = 5,000 AU from
the center is $n(H_2) \approx 2 \times 10^6$ cm$^{-3}$ and the
crossing-time is $t_{cross} \sim R / V <  5 \times 10^4$ years.
The most massive protostellar  objects are likely to either form in the 
center where the density and pressure are the greatest, or if they form 
away from the center, to rapidly migrate to the center of the potential 
due to orbit decay  (Zinnecker 1982; Bonnell et al. 2001).    

Orbits  can decay due to Bondi-Hoyle (BH) accretion as they move 
though clump gas and dynamic-friction caused by the gravitational pull of the
wake produced by converging streamlines behind the YSO.   
The BH accretion rate is roughly
$
\dot M_{BH} = \pi G^2 M_{YSO}^2 \rho(r) / (V_*^2 +  C_s^2)^{3/2}
$ 
where $C_s$ is the effective sound speed (which includes the effects of
turbulent motion and magnetic support)  in the gas (usually $< 1$ km~s$^{-1}$). 
A star moving with a speed  $V = 2.1$ km~s$^{-1}$ 5,000 AU from the 
center of the above  gas distribution  would accrete at a rate
$\dot M_{BH}  \approx  9  \times 10^{-7}$ M$_{\odot}$~yr$^{-1}$  for a
1 M$_{\odot}$ YSO and
$\dot M_{BH}  \approx  2 \times 10^{-5}$ M$_{\odot}$~yr$^{-1}$ for a 
5 M$_{\odot}$ YSO.  As illustrated by the simulation described in the
next paragraph, the most massive objects grow most rapidly,
doubling their  mass  and migrating to the
center of the cloud within $few \times 10^5$  years.  
Fragmentation  (McKee \& Tan 2002; 2003; 
Krumholz 2006)  or  the capture of sibling stars (Moeckel \& Bally 2006; 2007a,b) 
may produce multiple systems.     These  processes could have  
resulted in the assembly of a sub-cluster of massive stars and 
binaries in  the center  of the core.  

Figure 4 shows a sample result of a simplified numerical model illustrating the combined  
effects of Bondi-Hoyle accretion onto protostellar seeds,  orbit  decay, formation 
of a  non-hierarchical system of massive stars, and  dynamical 
ejection.Ê  This simulation starts with a 50 M$_{\odot}$ clump of gas in a Plummer potential 
given by $\Phi (r)  = -GM /  \sqrt{(r^2 + a^2)}$ where $M$ is the total mass and 
$a$ = 0.025 pc  is the core radius.    (This potential is used to avoid the singularity
at the center of an isothermal sphere). The clump is non-rotating with an internal 
velocity dispersion  (effective sound speed) given by the local Virial velocity so that, 
to first order, the clump is  stable to global gravitational collapse.   At the start of the 
simulation, 20 protostellar seeds,  each having an initial mass of 0.5 M$_{\odot}$, are 
distributed randomly in the clump.    Their  velocity distribution is virialized in the 
combined gravitational potential of the stars and  gas and their velocity vector orientations 
are random.  The simulation was run 100 times, each with a different randomly chosen 
set of initial locations and velocities for the seeds.

Stellar motions are calculated from the gravitational 
forces exerted by other stars and the static gas cloud, modified by the effects of BH 
accretion.  The inter-stellar forces are directly calculated, and the orbits are integrated 
with a global but variable time-step using a (7,8) order Runge-Kutta pair (Prince and 
Dormand 1981).  The code is a standard n-body integrator with the addition of a static 
potential, identified with the natal gas, that can be accreted onto protostellar seeds.  
The code follows the stellar mass growth and orbit evolution in the combined
gravitational potential of the clump plus embedded stars.  Similar approaches have 
been used by Moeckel and Clarke (2010) and Baumgardt and Klessen (2010). 
The protostellar seeds experience BH accretion from the clump and grow in mass 
with the accretion radius set to the minimum of the BH radius or one-third the distance 
to the closest neighboring star.  To keep  the computations simple, the response of 
the gas to the passage of the stars is not modeled.    This simplification is reasonable
because the zone influenced by the passage of a protostellar seed is restricted
to its gravitational radius, $r_G = G_* M / V_*^2$ ($ \sim $ 200 AU for 
$M_*$ = 1 M$_{\odot}$ and  $V_*$ = 2 km~s$^{-1}$).   Random motions in the 
clump will tend to fill-in the cavity formed by the passage of a star on  a time scale 
$\tau \sim r_G / C_s \ge$ $10^3$ years but much less than the $10^5$ 
year evolution time-scale of  the protostars and clump.  Thus, it is assumed that on 
average,   the envelope  remains spherical and smooth.  However, as gas
is accreted onto the stars the mass of the gas cloud,  and thus 
the contribution of the gas to the  gravitational potential, is reduced accordingly.   

Protostars  located in the densest  portion of the clump and/or those that by chance 
have the lowest relative velocity with respect to the  gas tend to grow in mass 
most rapidly.    In this simulation, accreted gas lowers the orbital angular momentum
of the stars about the clump center since the clump is assumed to be non rotating.   
Thus, as they grow in mass, their  
orbits decay and the most massive objects migrate to the clump center. 
The most rapidly growing protostars  form  a non-hierarchical system  near the center.
As they come to dominate the mass in  this region,  mutual  gravitational interactions 
reconfigure the system into a hierarchy consisting of one or more tightly bound compact 
binaries plus ejected high-velocity  stars.     Multiple runs of the simulation with 
different randomly chosen initial  locations and velocities almost always produce 
mass-seggregated,   non-hierarchical systems of massive stars at their centers.  
This model illustrates one plausible  scenario for the formation of a non-hierarchical 
system of massive stars and their  subsequent ejection.    Three essential features
are missing from this simplified model;  the dynamic ejection of gas from the central 
region by the orbital motion of  the massive stars, the presence of  accretion disks, and
the streamers that  transport mass from the inner-boundary of the envelope onto
these disks.  Previous simulations of binary star formation have shown these features
to be present (e.g. Lubow \& Artymowicz 1996).

In a multiple star system, matter that enters the region where the stars orbit each 
other tends to be expelled by gravitational torques.   As a result, gas 
bound to the  system is likely to be organized into two components; an extended 
outer envelope having an inner boundary with a radius somewhat larger than 
the semi-major axis of the largest stellar orbit, and inner disks surrounding 
the stars with outer radii  several times smaller than the periastron separations
(Artymowicz \& Lubow 1994; G{\"u}nther  \& Kley 2002).     
For pre-decay interstellar separations of 100 AU,   matter  located within  
$\sim$ 300 AU  of the cluster  would  either  be accreted onto  disks 
with outer radii  less than about 30 AU  or be expelled to beyond 300 AU.   
Gas in an envelope  bound to a 40 M$_{\odot}$ cluster of stars with an inner radius 
of 300 AU would have Kepler speeds less than 11 km~s$^{-1}$.     The outer 
boundary where the cluster has significant influence on the envelope can
be defined by the radius  where the gravitational  
influence of the cluster  falls below  the velocity dispersion,  
about 2 to 3 km~s$^{-1}$ for OMC1.  Disks with 30 AU outer radii
orbiting 10 M$_{\odot}$ stars would have Kepler speeds greater than 17 km~s$^{-1}$.
Since non-hierarchical  multiples likely have chaotic orbits, disks may be truncated 
at smaller radii.     Gas falling  from the outer envelope onto individual 
circumstellar disks may form  transient  streams in the region of avoidance.    

Figures 5 illustrates a possible scenario for the evolution
of the OMC1 cloud core.   Figure 5a shows the formation of several protostellar
seeds destined to become massive stars as they continue to accrete from the
clump and sink to its center  to form a non-hierarchical multiple system.   
Figure 5b shows the multiple system after it has cleared the orbit zone.   Tidal streams
can continue to feed the growth of compact circumstellar disks with outer radii
smaller than the typical periastron separation.  Figure 5c shows the aftermath
of the dynamic interaction that ejected the stars from the clump.  

\subsection{Recoil of the  Envelope; Slow Ejecta}

Proper motions   (Gomez et al. 2005, 2008) indicate that the stars 
BN, I, and n originated  from  a region 
less than 400 AU in diameter  500 years ago.   To evaluate the kinetic 
energy  of the envelope before stellar ejection,  consider a sphere of gas with
a central hole containing a  cluster with mass 
$M_{cl}$  = 20 to 40  M$_{\odot}$ (`cl' stands for cluster).      Assume
that the  envelope  density is given by  $\rho(r) = \rho_0 r^{- \alpha}$ between 
an inner boundary $r_{in}$ and an outer boundary  $r_{out}$.   
The initial total mass of the envelope may have been around
$M_e$ = 100 M$_{\odot}$.     As it formed  stars,  the 
envelope mass would decrease to  $M_e - M_{cl} ~\approx$ 60 to 80 M$_{\odot}$. 
Appendix 1 presents  an analytic formula for estimating the  kinetic energy
stored in the envelope due to orbital motion about the central cluster. 

Assuming an envelope  velocity dispersion of  $\sigma$ = 2 km~s$^{-1}$,   
the gravitational potential of the stars  dominated motions  located  within  
$r_G = GM_{cl} /  \sigma^2 \approx$ 4,400 to 9,000 AU (11\arcsec\  to 21\arcsec\  
at the 414 pc distance to OMC1) for 20 and 40 M$_{\odot}$ of stars, respectively.    
Clumps within the envelope at a distance $r$ from the center 
would move with velocities of order   $V(r) = (G[M_{cl} + M_{gas}(<r)]/r)^{1/2}$ 
=  3 to about 20 km~s$^{-1}$ in the  potential.    For a 0.1 pc outer radius 
sphere containing  60 M$_{\odot}$ of gas and  40 M$_{\odot}$ of stars, 
the mass of gas interior  to 5,000  AU is about 7  M$_{\odot}$ if the power-law index
of the density distribution  is $\alpha$ = 1.5, and almost 14 M$_{\odot}$ 
if $\alpha$ = 2.  Thus the total mass in this region is comparable to the low velocity 
portion of the BN/KL outflow.   

Following stellar ejection,   the inner envelope where the stellar mass
dominated the potential will be moving faster than the escape 
speed from the remaining mass.   Thus, this mass  will expand.
Assuming that each mass element moves with the Kepler
speed corresponding to its orbit radius before ejection, the average 
kinetic energy at radius $r$ in an infinitesimal radial increment $dr$ is  
$d \epsilon = 0.5 V^2_K(r) dM$ where $d M = 4 \pi r^2  \rho (r) dr$ and
$V^2_K(r) = G[M_{cl}+M_e(r)]/ r$, where $M_{cl}$ is the mass of the ejected stars
and $M_e(r)$ is the mass of envelope gas inside radius $r$.   This can be integrated
over radius to determine the total kinetic energy in orbital motion.  Following
stellar ejection, as this gas expands, it had to climb out of its own potential.
The free kinetic energy of the expanding envelope is given by the orbital 
kinetic energy before stellar ejection minus the energy required to climb 
out of its own potential well  after stellar ejection.

Appendix 1 gives an expression for the free kinetic energy of a recoiling
envelope which,  prior to stellar ejection, had mass $M_e$ with a 
power-law radial density profile between an outer radius $r_{out}$ and 
inner radius $r_{in}$ bound to  a central cluster with mass $M_{cl}$.     
Following stellar ejection, the envelope has more kinetic energy than 
(negative) gravitational potential due to its own mass.    
The orbital kinetic  energy due to the central  cluster is  represented by the 
first term in equation A1-1.     The self-energy of the envelope is represented 
by the second term in equation A1-1.    Following dynamic ejection, the 
orbital motion of clumps in the envelope  exceed the self-binding energy.  
Equation A1-2 gives the final kinetic  energy of the envelope  after ejection 
of the central cluster.  Table 1 lists  these kinetic energies  
($10^{45}$ to $10^{46}$ ergs)  for  parameters  considered plausible for OMC1.

For steeper density gradients,  a larger fraction of the envelope is concentrated 
deeper in the potential well of the stars.   The  kinetic energy of the ejected 
envelope  can be increased by migrating mass closer to the center of mass 
prior to stellar ejection.   For example, if the entire  mass of an inner envelope
with M$_e$ = 10 M$_{\odot}$  were 
placed  in an orbiting ring with  a radius $r$, its orbital kinetic  energy 
would be  $GM_{cl} M_e /  2 r$ $\approx 3.5 \times 10^{46}$ ergs 
for $r$ = 100 AU, $M_{cl}$ = 40 M$_{\odot}$ .
The orbital momentum of such a ring, $P_e = M_e V_K(r_{in})$ would be  about 
190~M$_{\odot}$~km~s$^{-1}$, comparable to the outward directed momentum in
the BNKL outflow.   The gravitational self energy of the inner envelope is
of order  $GM_e^2/r \approx 1.8 \times 10^{46}$ ergs (the exact number
depends on the density and dimensions of the ring).  Thus, in this example,
the outflowing kinetic  energy of the envelope after stellar ejection is about
one half of the orbital kinetic energy.   The fraction of the orbital energy  available 
as free kinetic energy in an outflow increases as the ratio of the envelope 
mass to  the ejected cluster mass decreases.

\subsection{Disruption of Disks; Fast Ejecta}

The destruction and ejection of circumstellar matter during the final penetrating 
encounter leading to the decay of a massive multiple star system that ejects 
several members can release an order-of-magnitude more gravitational 
potential energy than the gentle recoil of  the surrounding weakly bound 
envelope.  During the final stellar encounter that formed an  AU-scale binary  
whose  gravitational potential energy expelled the stars  from the region,  their  
circumstellar disks would likely be destroyed and ejected.     If  the binary 
consists of a pair of 10 M$_{\odot}$ stars separated by 0.7 to 2.2 AU as 
required by the energetics of the OMC1 ejection, the outer radius of any 
surviving disk around either star can be no greater than about 0.1 to 0.5 AU.   
Matter ejected from this region in orbit around a 10 M$_{\odot}$ star would 
have a velocity of order 100 to 300 km~s$^{-1}$.   In an encounter between 
two or three massive stars, gravity will accelerate the stars to about this speed
if the periastron is about an AU.  In a prograde encounter between a star  and
matter in another star's disk,  the head-on flow  can ``slingshot"  this material
to about twice the relative encounter speed.    Moeckel and Bally (2006, 2007a, b)
presented SPH simulations of the gravitational and hydrodynamic interaction 
of single stars crashing through the disk of a massive star on hyperbolic orbits.  
They found that for a wide range of parameters, the incoming stars could be 
captured into an eccentric, non-coplanar, elliptical orbit.   The interaction 
truncates the disk, and ejects its outer parts.

The fastest observed ejecta in the BN/KL outflow have speeds between 
200 to 400  km~s$^{-1}$ but the total mass in the fast components
above 100 km~s$^{-1}$ is less than 0.1 M$_{\odot}$.   We propose that 
this ejecta was produced by the disruption of circumstellar disks
and launched from radii   $r_d <  $ 1 AU from around  $\sim$ 10 M$_{\odot}$  
stars.      Assume that  individual circumstellar disks can be characterized by
a power-law surface density of the form $\Sigma_d = \Sigma_{0} r^{-\beta}$
where the subscript $d$ signifies  the inner disk (as opposed to the extended 
envelopes).       Appendix 2 gives an expression for the free kinetic energy 
of  high-velocity `shrapnel'  created by the disruption of a circumstellar disk 
during the final encounter that led to the formation of a compact binary.   
Table~2 lists the kinetic energies of this debris as a function of plausible
pre-ejection disk parameters.        For  power-law disks with  
$\beta$ ranging from 0.5 to 1.5,   outer radii $r_{d\_out} = 10$ AU, 
inner radii ranging from $r_{d\_in}$ =  0.05 to 0.5  AU,   total disk masses  of 
0.5 to 1.0 M$_{\odot}$ orbiting  a $M_*$ = 10 M$_{\odot}$ star,    
the kinetic energies of ejecta after stellar ejection range from
$10^{46}$ to $10^{47}$ ergs, an order of magnitude greater than the energies in
the recoiling envelope.   Steeper surface density profiles  concentrate 
matter toward the star and  increase the kinetic energy per unit mass stored 
in the disk.   Table~2 lists the energies of ejected disk gas.     If the  circumstellar 
material  surrounding several stars are ejected 
in this manner,  the kinetic energy and  momentum requirements of the BNKL 
outflow can  be met by the disruption of such  disks alone.   The disruption and 
ejection  of the inner disks during the final  encounter  that ejected OMC1's 
massive stars could have powered the fastest ejecta  in the BN/KL outflow.

\subsection{The Contribution of Stored  Magnetic Stress}

The birth and subsequent  orbital motion  of a  non-hierarchical cluster of 4 or 5 
massive stars  in OMC1 over $10^5$ or $10^6$ years may have led  to the 
amplification of ambient magnetic fields in both the envelope surrounding the
cluster and in individual circumstellar disks.   
Fields can grow  until either  magnetic-reconnection or resistive dissipation 
in the charged  component of the weakly ionized plasma come to equilibrium 
with the dynamo  (Lynden-Bell 1996;   Wheeler, Meier, \& Wilson 2002; 
Vasil \& Brummell 2008, 2009). 
The orbital motion of the protostars in the cluster gravitational 
potential and the orbital motion of gas in individual circumstellar 
disks may each contribute to the amplification of the magnetic fields 
(K{\"a}pyl{\"a} et   al. 2008).  
 
When equipartition is  reached, $B^2 \approx  8 \pi \rho (r)  c_s^2$,  
where $\rho (r)$  is the local gas density  and $c_s$ is the effective sound 
speed including turbulent motions.    For the envelope having the
power-law density profile described in Appendix 1, the equipartition
magnetic energy density at radius $r$ is given by  
$\epsilon _{Be}  = B^2 (r) / 8 \pi = \rho (r) c^2 _s$ where $c_s$ is the local effective
sound speed (including thermal and turbulent motion) and the subscript
$e$ refers to the envelope.    Under this assumption, the magnetic energy
in a shell at radius $r$ and thickness $dr$ is  
$$
dE_{Be} = 4 \pi r^2 \epsilon _{Be}  dr
           = r^2 B_e^2(r) dr / 2  = \rho _e(r) c^2_s dr
           = 4 \pi \rho _0 c^2_sr^{2 - \alpha} dr
$$
which can be integrated from the inner edge of the envelope
to a radius $r$ to give
$$
E_{Be} = \int _{r_{in}} ^{r}   dE_{Be} 
 = {{4 \pi \rho _0 c^2 _s} \over {3 - \alpha}} ( r ^{3 - \alpha} - r_{in} ^{3 - \alpha}).
$$
Setting $r = r_{out}$ gives $E_{Be} = M_e c_s^2$.  For 
$M_e$ = 60 M$_{\odot}$ and $c_s$ = 2 km~s$^{-1}$ this results in
$E_{Be} \approx 5 \times 10^{45}$ ergs for the equipartition
magnetic energy that could be stored in the envelope.  

If magnetic stress reaches equipartition with the effective sound speed within
circumstellar disks, a similar calculation yields $E_{Bd} = M_d c_d^2$ where
the subscript $d$ refers to the disk quantities.   The effective sound speeds
in the disks are likely to be considerably larger due to proximity to the massive stars.
Using $c_s$ = 10 km~s$^{-1}$ gives  $E_{Bd} = 2.0 \times 10^{45} M_1 c^2_{10}$
ergs for the equipartition magnetic energy that could be stored in a disk.    
Here $M_1$ is in units of 1 M$_{\odot}$ and $c_{10}$ is in units of 10 km~s$^{-1}$.

Magnetic fields may become stronger than  equipartition with the local 
pressure in the disk or envelope.      Keplerian shear  can wind-up 
the field so that its energy approaches the gravitational potential,
a circumstance  discussed in the context of circumstellar disks by  
Shu et al. (2008).   The condition that  Alfven speed (or effective sound 
speed) is  equal to the  gravitational escape  speed sets an  upper bound 
on the field-strength.   Thus, $V^2_A  = B^2 / 4 \pi \rho \approx  GM/r$ 
where  $M$ is the mass enclosed  inside radius $r$.   
In this case, the field could have stored  more than  $10^{47}$  ergs of 
magnetic energy in the volume inhabited  by the cluster.   The removal 
of about  40 M$_{\odot}$ of stars would have  left the  circum-cluster gas  
severely magnetically   over-pressured.  The release  of magnetic energy,   
in effect a   `magnetic bomb'  (Matt, Frank, \&  Blackman 2004, 2006), may 
have  been  a major source of energy  for launching the OMC1 outflow.   
The strongest fields are likely to be generated by shear dynamos in 
the innermost parts of circumstellar disks surrounding individual massive 
stars.  Future  Zeeman measurements of clumps containing sub-clusters of
massive protostars using the 24 to 26 $\mu$m iron lines 
with SOFIA may determine if energetically significant, gauss-strength 
fields can be generated in such environments 
(R. Crutcher - private communication).

\section{Discussion}

The BN/KL outflow is velocity segregated with relatively slow gas
confined to a few tens of arc-seconds from radio source I and the fastest ejecta, 
located near the tips of the [FeII] and H$_2$ fingers  about 120\arcsec\
northwest and southeast of the core region.     Proper motion measurements of 
visual-wavelength HH objects (such as HH 205 through 210), near-infrared
emission from shock-excited [FeII] and H$_2$ (Figures 1 and 2), and  the interferometric 
CO J=2--1 study of Zapata et al. (2010) indicate that the BN/KL
outflow was generated by an explosive event about 500 years ago.   
This age coincides with the dynamical re-arrangement of the massive 
stars in OMC1 during which radio sources I, n, and BN were ejected  
(Gomez et al. 2005; 2008).     The slowest star, radio source I,  is probably 
a compact binary consisting of roughly 10 M$_{\odot}$ stars with a separation 
less than about 2 AU.   Future long-baseline radio interferometry,  or precision 
radial velocity measurements in the infrared may determine if this conjecture is true.   
It is proposed  that the formation of this binary and the consequent 
release of $2$ to $6 \times 10^{47}$ ergs of gravitational potential energy  
powered both the motion of the stars and the supersonic expulsion of gas 
in the BN/KL outflow.   The slowest ejecta may  have been produced at 
large radii as the  pre-ejection orbits of clumps in the 
envelope became unbound following stellar ejection.    
The fastest  ejecta may have originated from near the center of the 
potential well vacated  by the  stars.  This situation is similar to that postulated 
by McCaughrean and Mac Low (1997) and Stone et al. (1995),  where the H$_2$ 
fingers are generated by  the interaction between a slow, outer wind and a fast, 
later ejected flow.     The estimates  based on simple models  of  envelopes and 
circumstellar disks with power-law density distributions indicate 
that the gravitational energy stored in orbital motion prior to ejection can 
explain the energetics of the outflow following stellar ejection.   However, 
numerical modeling of realistic configurations is need to determine of this 
hypothesis is valid.

The explosive OMC1 outflow consists of multiple fingers and wakes of H$_2$ 
emission produced by the fastest ejecta plowing through the OMC1 core.
The volume filling  factor of  the cavities  created by this ejecta  is likely to 
be lower than the enclosed volume,  especially  towards the southeast.    
Observations show that  a reservoir of dense gas  is located in the `hot core'   
5\arcsec\  to 20\arcsec\   southeast of the site of  stellar ejection.   
Zapata et al. (2010)  propose that  the `hot-core'  may be blocking portions 
of the outflow in this direction.    The  bulk of the H$_2$  emission 
is located to the northwest,   consistent with  more of the original  core  
being blown-out in this direction.    The ejected  stars moving south  likely 
plow through the remaining dense material.

\subsection{ Post-Ejection Disks and Outflow Orientations} 

Source I  and possibly  BN and  n,  are surrounded by circumstellar disks 
(Rodriguez,  Zapata, \& Ho 2008).    A polarized, infrared bipolar reflection 
nebula can be traced over 20\arcsec\  from BN.  Simpson et al. (2006)
argue that the polarization is produced by dichroic absorption
by magnetically aligned dust in a foreground  dust lane.    Jiang et al. (2005) 
presented-adaptive optics-assisted polarimetric near-infrared images of  the 
BN object and found  a  polarization pattern indicating illumination from BN.   
The nebula is bipolar  with a symmetry  axis  oriented at 
PA $\approx$ 36\arcdeg\  and a  dark band parallel to BN's proper motion. 
The dark lane may be a ``disk-shadow" (e.g. Pontoppidan \&
Dullemond 2005)  produced by opaque material close to the central star;  thus
the silhouette only provides an upper-bound on the disk outer radius.  
Such a disk must be oriented close to edge-on with an axis nearly orthogonal 
to BN's  proper motion.    

Radio source I is  also surrounded by a nearly edge-on disk 
with an axis orthogonal to its  apparent proper motion.  The elongated 7 mm
continuum emission associated with source I  has been  interpreted as  a 
nearly edge-on disk rendered visible by  collisional ionization and H$^-$ 
free-free opacity (Greenhill et al.  1998; Reid et al. 2007).    The disk has a 
radius of about 50 AU and an axis oriented  towards  PA $\approx$ 45\arcdeg .    
The thermal SiO emission at  86 GHz (Plambeck et al. 2009)
and  22 GHz H$_2$O  masers on scales of a few thousand AU indicate the 
presence of a very compact  bipolar outflow with an age of only 
a few hundred years blowing towards the northeast and southwest along the 
suspected disk axis (Wright et al. 1995; Matthews et al. 2008; 
Beuther \& Nissen 2008; Plambeck et al. 2009).

Although source n is suspected to be surrounded by a disk, 
its orientation remains uncertain.  Elongation in mid-infrared images
suggest an axis towards the north-northeast, close to the axis of the
elongation or double structure seen at radio wavelengths
(Shuping et al. (2004). 

Acceleration of the  stars during the interaction that ejected them 
would have stripped away the outer parts of  disks where the Kepler 
speed in less than than the stellar velocity.    Thus, for BN and n,  surviving 
disks would be truncated beyond about  10 AU. The final penetrating 
encounter during the decay of the  non-hierarchical multiple star 
system may place even more stringent constraints on disk outer
radii by ejecting 
material lying outside a radius of around 1/2 to 1/3  the periastron 
distance of the encounter.      If source I consists of a pair of
10 M$_{\odot}$ stars,  material beyond about an AU  would
be ejected.    Some of this material may have remained bound to source I 
and fallen back to form a circum-binary  disk.   Such a disk would be 
confined to an annular region with a radius greater than 3 to 10 AU set 
by the binary orbits,  and less than about 50 AU by the ejection velocity.   
The 7 mm continuum and SiO maser emission from source I indicates 
that it is currently surrounded by a disk with an outer radius of about 
50 AU with a northeast-southwest axis.  A very young (dynamic age of
a few hundred years) outflow emerges  along this axis for a few 
arc-seconds.     Did the source I disk survive  dynamic ejection, or 
was it accumulated from the surrounding core within the last 500 years?
These questions are considered in the following sub-section.

\subsection{Disks Accreted from the Surrounding Medium?}

In the proposed scenario, the currently observed disks 
either  trace 
(i) material that has expanded from within a few AU of each star,   
(ii) disk debris ejected at less than escape speed that
fell-back to re-form a new disk in the last 500 years,  or 
(iii) was captured by  Bondi-Hoyle accretion as the ejected stars plowed 
through remaining dense gas such as the `hot core' in OMC1. 

A star with mass $M_*$ moving with velocity $V$ through a cloud 
can accrete material lying within a cylinder defined by the  gravitational 
radius   $r_G = GM_* / V^2$ which is measured orthogonal to the 
stellar trajectory.    
After traversing a distance $L$ the amount  of mass captured is about
$$
M_{BH} = \pi r_G^2 L  \rho =  \pi \mu  m_H n(H_2) L G^2 M_*^2 / V^4
= 2 \times 10^{-5} n_5 L_3 M^2_{10} V^{-4}_{10} ~~~~~~(\rm M_{\odot})
$$
where $\mu$ is the mean molecular weight, 
$m_H$ is the mass of hydrogen,
$n_5$ is the number density of H$_2$ in units of $10^5$ cm$^{-3}$,
$L_3$ is the path traversed by the star in units of $10^3$ AU,
$M_{10}$ is the mass of the star in units of 10 M$_{\odot}$, and
$V_{10}$ is the velocity of the star through the medium in units
of 10 km~s$^{-1}$. 
Throop and Bally (2008)  explored some consequences of BH accretion 
onto low-mass protostars as  they moved through dense clouds.  
Moeckel and Throop (2009) modeled this late-phase BH accretion 
from a uniform density cloud onto a young star
surrounded by an accretion disk using a hydrodynamic code with an
isothermal equation of state  but did not consider velocity or density
gradients.      

Bondi-Hoyle accretion  tends to  be subject to a ``flip-flop'' instability 
that leads to highly variable  accretion  rates and fluctuating  orientations 
of the angular momentum  vector of the flow (Shima et al. 1998; Fryxell \& 
Taam 1988; Ruffert 1987, 1999;  Krumholz, McKee, \& Klein 2006).    
However, gradients in either  velocity  or density orthogonal  to the stellar 
velocity  give the flow vorticity with respect to the accretor.    The resulting 
angular  momentum  can curtail  accretion directly onto the moving mass
because the flow tends to form a disk-like structure.    Ruffert (1997)
presented 3D simulations of Bondi-Hoyle accretion from a medium having
a velocity gradient while Ruffert (1999) considered flows from media
with density gradients.     Krumholz,  McKee, \&  Klein (2005) modeled accretion 
from a medium having vorticity.     These studies show that Bondi-Hoyle accretion 
from a medium with a small velocity or density gradient is still subject to
the ``flip-flop" instability.  However,  large gradients tend to produce 
accretion disks with a relatively stable  angular momentum vector orthogonal  
to the star's velocity (Ruffert 1999).

The BN object is moving away from the hot-core towards the northwest and
has traversed about 4,000 AU since ejection.  If the average density of the 
medium through which it moved had $n(H_2) \sim$ 10$^5$~cm$^{-3}$, a 30
km~s$^{-1}$ velocity implies that it could have swept-up about   
$M_{BH}$ = 1.4 $\times$ 10$^{-6}$ M$_{\odot}$.    If source n has a mass
of 5 M$_{\odot}$, traversed 2,000 AU through a medium with a density of 
10$^{5}$~cm$^{-3}$ with a velocity of 30 km~s$^{-1}$,  it could have 
Bondi-Hoyle accreted $M_{BH}$ = 1.2 $\times$ 10$^{-7}$ M$_{\odot}$. 

Radio source I is moving towards the dense `hot core'.  Assuming that it has a 
mass 20 M$_{\odot}$, is moving through a medium with  
$n(H_2)$~=~ 10$^6$~cm$^{-3}$ with  a velocity of 10 km~s$^{-1}$,
and has traversed a distance or order 1,000 AU, it could have swept-up as
much as    $M_{BH} = 8 \times 10^{-4}$ M$_{\odot}$.    
For source I, assuming a mass $M_I$ = 20 M$_{\odot}$,
the time to cross the gravitational radius, $GM_I/V_I^2 \sim$~100~AU 
at velocity $V_I$ = 13 km~s$^{-1}$ is less 
than 40 years.    The Kepler orbit time at a distance of 50 AU from a 
20 M$_{\odot}$ compact binary is about 80 years.  Thus, any material 
accreted within the last several hundred years would have orbited at least 
three  times at 50 AU.  Thus, enough time has elapsed for  accreted matter 
to have damped much of its motion orthogonal to its plane of symmetry and 
would resemble a disk.    

Both BN and radio source I are moving  close to the plane of the sky.   If the 
accreted disk's angular momentum vector lies in a plane orthogonal to
the stellar velocity vector, the apparent major axis of such a disk as projected
onto the plane of the sky will be aligned with the proper motion vector.
Observations show that the major axes of both the 7 mm continuum 
emission (Reid et. al. 2007) and the SiO maser emission
(Goddi et al. 2009) from radio source I  (which are interpreted as a nearly 
edge-on disk), and the putative disk  associated  with BN (Jiang et al. 2005) 
are  parallel to their proper motions.     These disk orientations are consistent
with Bondi-Hoyle accretion onto the ejected stars from a medium having a 
significant density or velocity gradient orthogonal to the stellar motion.

\subsection{Is the Orion BN/KL Outflow Unique?}

If such a mechanism operates in other hot cores and massive star forming 
regions, then it is expected that observations will reveal other outflow systems
similar to Orion BN/KL.  
Such eruptive or explosive outflows will be associated with ejected, 
high-velocity massive stars whose time since ejection will be comparable
to the outflow ages.    The Spitzer Space 
Telescope detected 4.5 $\mu$m emission from a wide-angle outflow having a
morphology similar to Orion emerging from  a high-luminosity 
($\sim 10^6$ L$_{\odot}$) hot-core in  G34.25+0.16 in the inner Galaxy
(Cyganowski et al. 2008).     Unfortunately, this flow and its source cloud 
core are highly obscured  because it is located at a distance of about 5 kpc 
in the Galactic Molecular Ring.    Another possible example of an outflow 
having a morphology suggestive of an explosive origin is located in
the NGC 7129 star forming region in Cepheus.     This flow appears to
originate from a moderate luminosity ($< 10^3$ L$_{\odot}$) protostar.  
Source G in W49,  which is the most luminous water maser outflow in the 
Milky Way, may be yet another example (Smith et al. 2009).
Finally Sahai et al (2008) found evidence for interstellar bullets having
a similar structure to the OMC1 fingers in the outflow from the
massive young protostar  IRAS 05506+2414.    
  
\section{Conclusions}

The main conclusions  of this study are:

1) New proper motion measurements show that some of the fastest
and most distant H$_2$ knots in the OMC1  BN/KL outflow have 
dynamical ages of  500 years, similar to the 500 year interval since 
dynamical decay ejected radio sources I, BN, and n from a region 
located only a few arc-seconds from  the center of the outflow.   
Thus, it is likely that the outflow was produced by the same 
event than ejected the massive  stars BN, I, and n  form the core.

2) To within an order of magnitude, the kinetic energies of the outflow and 
ejected stars are  comparable.  

3) A scenario is explored in which the explosive OMC1 outflow is produced
by the dynamical ejection of massive stars.   Within the last few hundred 
thousand years, a  compact non-hierarchical multiple system of massive 
stars or a pair of massive binaries must have formed in  the OMC1 cloud core.   
A highly idealized numerical model is presented in which protostellar seeds
grow by Bondi-Hoyle accretion from the parent clump.   Seeds which 
by chance wander into the densest part of the core or have the lowest relative
velocity with respect to the surrounding gas experience the highest accretion
rates.   They become massive,  experience orbit decay,  and migrate to the 
center of the model where they form a dynamically unstable multiple 
system.  

4)  The kinetic  energy of the stars and the outflow could have been 
generated by release of gravitational binding energy accompanying the 
formation of a compact binary, most likely radio source I.   If  source I consists 
of a pair of 10 M$_{\odot}$ stars, then to produce the observed kinetic
energy of the ejected stars and the outflow, between $2$ to $6 \times 10^{46}$
ergs, the mean separation of the binary must be between 2.2 and 0.7 AU. 

5)  In the proposed scenario, the outflow is driven by the energy and 
momentum stored in orbital motion prior to dynamical decay.    
Previous studies of gas flows in multiple-star systems
show that orbiting material tends to be organized in a hierarchy
consisting of tightly bound circumstellar disks with outer radii less than 
about 1/3 of the periastron separation and relatively loosely bound 
envelopes with inner radii several times the apastron separation. 
Such a scenario is envisaged to to have been present in OMC1
prior to the ejection of its massive stars.

6)  Stellar ejection may contribute to the formation of an explosive outflow
by three mechanisms:
{\it Recoil of the envelope:}  The removal of stellar mass from the center 
of the envelope would result in the conversion of its orbital motion into 
linear motion.  
{\it   Disruption  and ejection of  circumstellar disks:} 
The final penetrating encounter  that formed a compact binary would eject
the inner parts of pre-existing disks to produce the fastest ejecta.
{\it Release of magnetic stress:}  
Magnetic energy potentially produced by shear-dynamo action in both
circumstellar disks and the envelope might boost the velocities of the ejecta.
In this scenario, the fastest ejecta is  launched first from deep inside
the gravitational potential of the decaying cluster.   This material must plow
through the slower-moving and later ejected envelope.   Such  interactions 
are prone to Rayleigh-Taylor  type instabilities as shown by models
of fast-winds slamming into slower, previously ejected winds.    
Order-of-magnitude energy estimates show that all three
mechanisms are plausible,  and that all three may contribute to the 
energy budget.

7) The ejected high-velocity stars may have accreted new circumstellar material
as they traversed the dense OMC1 core.   In the presence of  gradients in 
the medium, the angular momentum vectors  of accreting gas tend to be
orthogonal to the stellar velocity vectors.    Therefore, for stellar motions close to 
the plane of the sky,  disk  major axes will be  aligned (in projection) 
with the stellar proper motion vectors, as observed for both radio source I and BN.
Consequently, any recent outflow activity produced by the ejected stars  will have axes 
orthogonal to the stellar proper motion vector.   The youngest  ejecta from 
radio source I as traced by H$_2$O and SiO masers and CO 
has an axis orthogonal to its proper motion vector.  The axis of symmetry of the 
near-infrared bipolar reflection nebula associated with BN is also
orthogonal to its proper motion.

Future  numerical hydrodynamic or magneto-hydrodynamic modeling of a 
massive-star forming clump in which a non-hierarchical system experiences a
dynamical decay are needed to determine if the proposed scenario can indeed
result in the generation of a powerful, explosive outflow such as is observed in 
Orion.

\acknowledgments{
This work was supported by NSF grant 
AST0407356 and the CU Center for Astrobiology funded by NASA under 
Cooperative Agreement no. NNA04CC11A issued  by the Office of Space 
Science.  This paper is partially based on observations obtained with the 
Apache Point  Observatory 3.5-meter telescope, which is owned and 
operated by the Astrophysical  Research Consortium.  We thank
Bruce Elmegreen,  Hans Zinnecker, and Bo Reipurth  for  their insights 
and comments.   We thank the referee, Paul Ho, for a thorough reading and
very helpful criticism that greatly improved the manuscript.}

\clearpage

\section{Appendix 1:  Excess Kinetic Energy of a Power-law Envelope After Ejection 
of an Interior Cluster.}

Assume that a gaseous envelope surrounds a compact cluster of stars 
with mass $M_{cl}$, and that the motion of the gas is dominated by 
gravitational forces.  The stars are all interior to the inner spherical 
boundary at radius $r_{in}$, and the average gas density is given 
by a power law so that $\rho(r)  = \rho_0 r^{-\alpha}$ out to an exterior 
radius $r_{out}$.  If the total mass of the envelope is $M_e$, the 
normalization constant is 
$$
\rho_0 = {{(3-\alpha)M_e}  \over {4\pi (r_{out}^{3-\alpha} - r_{in}^{3-\alpha})}}
$$
The potential energy of the envelope is found from 
\begin{align}
W_e &=-4\pi G \int_{0}^{\infty} r \rho(r) M_{tot}(r) dr \notag\\
&=  -4 \pi G \int_{r_{in}}^{r_{out}}  \rho_0 r^{1-\alpha} M_{cl} {\rm d}r 
       -4 \pi G \int_{r_{in}}^{r_{out}}  \rho_0 r^{1-\alpha} M_e(r) {\rm d}r
\nonumber{~~~~~~~~~~~~~~~~~~~~~~~~(A1-1).}
\label{PotentialEnergyEnvelope}
\end{align}
where $M_{tot} = M_{cl} + M_e(r)$ is the total mass inside radius $r$ before
ejection of the cluster of massive stars.
The first term is associated with the central cluster mass, while the 
second term is the potential associated with the gas envelope itself
$M_e(r)$ being the envelope mass inside radius (r). 
Taking the pre-stellar-ejection envelope to be virialized, we have 
$-W_{init} = 2T_{init}$.  Because the  observed stellar ejection velocities 
are  much faster than the Kepler velocities in the envelope,  stellar 
ejection  happens on a timescale fast compared to the speed at 
which the gas can redistribute itself.   
Immediately after ejection of the stars,  the potential energy of the 
envelope will be given solely by the second term, which remains 
the same, and the kinetic energy will still be given by $T_{init}$.  
If the post-ejection envelope were virialized, it would have 
$-W_{final} = 2T_{final,vir}$; the excess kinetic energy is then given by 
\begin{equation}
T_{excess} = T_{init} - T_{final,vir} = (-W_{init} + W_{final})/2 = | \Delta W | /2.
\nonumber
\end{equation}
The quantity $| \Delta W |$ is simply the term in equation (A1-1)
associated with the cluster mass, which integrates to give
\begin{equation}
T_{excess} = 
      \frac{G M_{cl} M_e}{2} \frac{ (3 - \alpha)}{(2 - \alpha)} 
      \left [ 
      \frac{r_{out}^{2-\alpha} - r_{in}^{2-\alpha}}
      {r_{out}^{3-\alpha} - r_{in}^{3-\alpha}}
       \right ]
\nonumber{~~~~~~~~~~~~~~~~~~~~~~~~~~~~~~~~~~~(A1-2).}
\end{equation}

\clearpage

\section{Appendix 2:  The Stored Kinetic Energy of an Idealized  Power-law Disk}

To estimate the kinetic energy stored in the gravitationally bound motions of a
circumstellar disks surrounding a star of mass $M_{\star}$, assume
that between the inner edge
of the disk, $r_{d,in}$, and the outer edge at radius $r_{d,out}$, the
average gas surface density is
represented by a power-law on the radius of the form
$\Sigma(r)=\Sigma_0r^{-\beta}$.  With a total disk mass $M_d$, the
normalization is given by
 \begin{displaymath}
 \Sigma_0=\frac{(2-\beta)M_d}{2 \pi( r_{d,out}^{2-\beta}-r_{d,in}^{2-\beta})}.
 \nonumber{~~~~~~~~~~~~~~~~~~~~~~~~~~~~~~~~~~~~~~(A2-1)}
\end{displaymath}

Assuming a disk in near-Keplerian rotation (thus with a disk mass
small compared to the stellar mass), the potential energy of the
system is to a good approximation just due to the stellar potential,
\begin{equation}
W_d \approx -2 \pi G \int_{r_{d,in}}^{r_{d,out}}
\Sigma_0r^{-\beta}M_{\star}{\rm d}r
 \nonumber{~~~~~~~~~~~~~~~~~~~~~~~~~~~~~~~(A2-2)}
\label{PotentialEnergyDisk}
\end{equation}
With Keplerian motion we have $-W_{d,init} = 2 T_{d,init}$.
Immediately after ejection of the star, which we assume happens on a
timescale fast compared to the gas redistribution timescale, the
potential energy of the disk will be due only to its self-potential,
which we have neglected due to its assumed small contribution to the
total.  The excess kinetic energy is thus
\begin{displaymath}
T_{d,excess} = T_{d,init} - T_{d,final,vir} = (-W_{d,init} +
W_{d,final})/2 = | \Delta W_d | /2.
\end{displaymath}
Since the stellar potential was dominant, the quantity $| \Delta W_d
|$ is simply the potential energy due to the star from equation
\ref{PotentialEnergyDisk}, which integrates out to
\begin{equation}
T_{d,excess} = \frac{G M_{\star} M_d}{2} \frac{ (2 - \beta)}{(1 -
\beta)} \left[\frac{r_{d,out}^{1-\beta} -
r_{d,in}^{1-\beta}}{r_{d,out}^{2-\beta} - r_{d,in}^{2-\beta}}\right]
\nonumber{~~~~~~~~~~~~~~~~~~~~~~~~~~(A2-3).}
\end{equation}

\clearpage

\begin{figure}
\center{  \includegraphics[angle=0,scale=0.9]{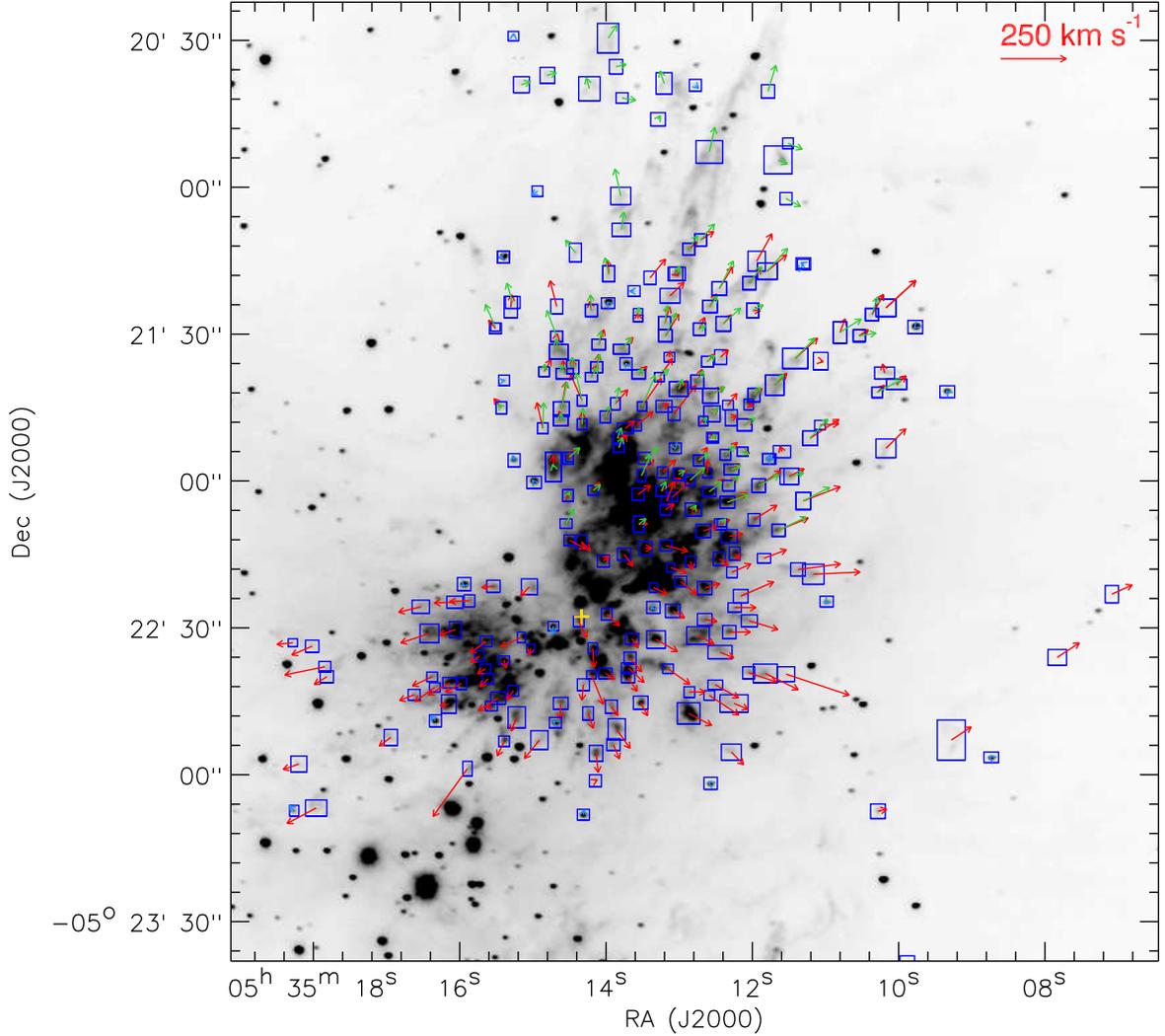}}
\caption{BN/KL outflow imaged in H$_2$ 2.12 $\mu$m emission in December 
2004,  from Cunningham (2006).   Blue boxes indicate features with proper motion 
measurements.  Green arrows denote proper motion measurements based 
on comparison of  the 2004 data with prior-epoch images from 
September 1992 (Allen \& Burton 1993).  Red arrows denote motions measured using
January 1999 Subaru images (Kaifu et al. 2000).   Short blue arrows 
indicate proper motion measurements on stars 
in the field, and verify proper image registration.  Arrow lengths indicate velocities 
according to the scale at upper right, and are equivalent to 122-year motions. 
The yellow cross indicates the apparent point of origin of three ejected, massive 
stars, radio sources BN, I, and n  (Gomez et al. 2005, 2008).}
\end{figure}
\clearpage

\begin{figure}
\center{  \includegraphics[angle=0,scale=0.9]{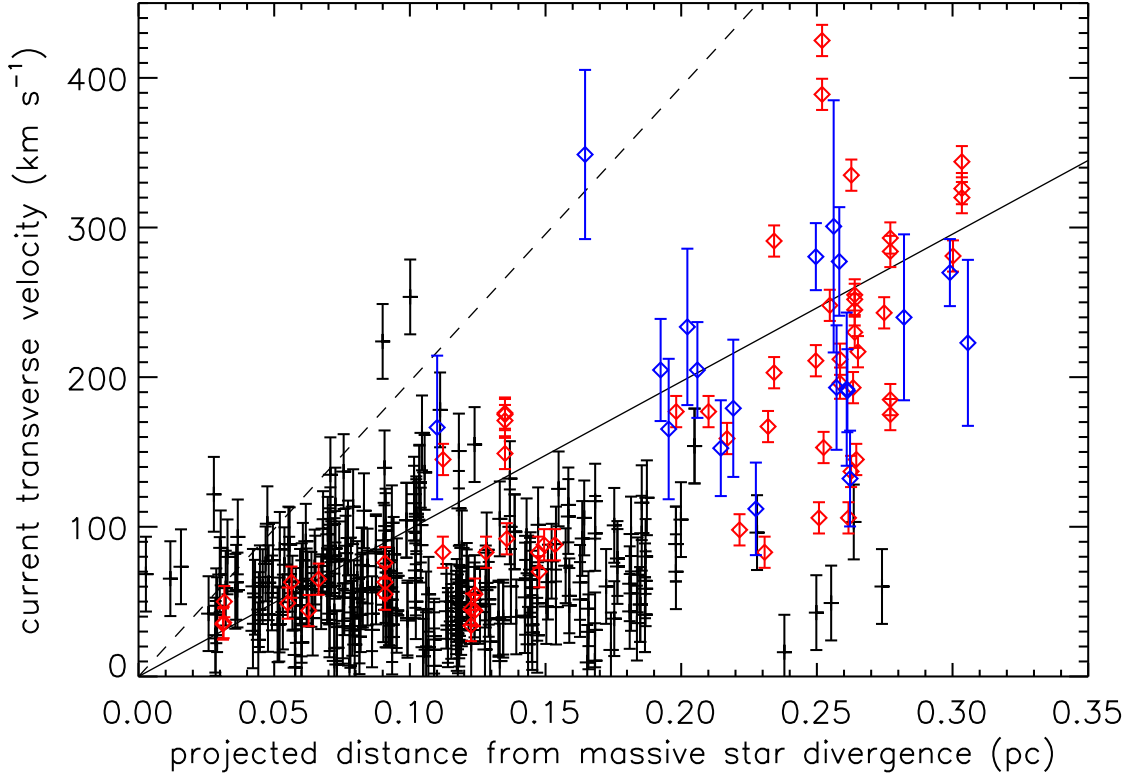}}
\caption{Projected distance versus transverse velocity for features in BN/KL with 
measured proper motions.  Distances are measured relative to the center of 
divergence of the massive stars as reported by Gomez et al.  (2005, 2008).  Features 
and proper motions are optical HH objects measured by Doi, O'Dell, \& Hartigan 
(2002, red  diamonds),   [Fe II] 1.64 micron bullets measured by Lee \& Burton 
(2000, blue  diamonds), and H$_2$ 2.12 $\mu$m features measured by 
Cunningham (2006,  black crosses).  The solid line indicates a zero-deceleration 
track for features launched 1000 years ago from the center of the dissociated system, 
and the dashed line indicates 500 year ages with no deceleration.  The presence 
of many data points above the 1000 year track, together with the expectation of 
deceleration for these knots as they lose momentum and energy via interactions 
with the ambient medium, suggests an age younger than the typically-reported 
$\sim$ 1000 years, and more consistent with the 500 year age of the multiple 
system dissociation event.}
\end{figure}
\clearpage

\begin{figure}
\center{  \includegraphics[angle=0,scale=1.4]{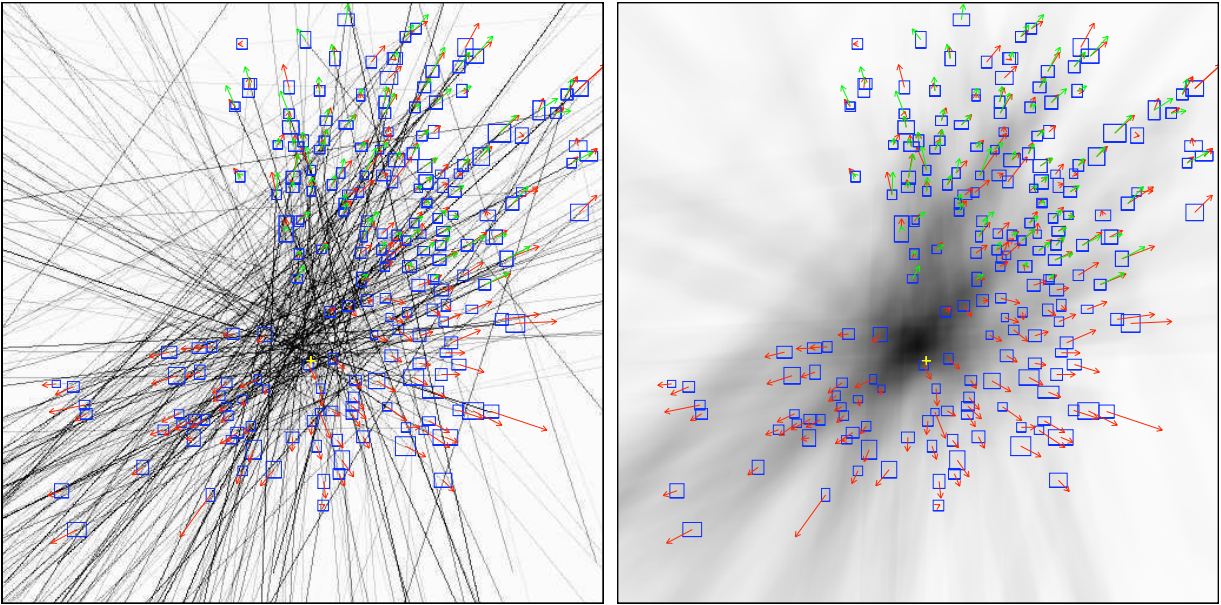}}
\caption{
Results of  the ``trail" analysis  to determine the point of origin of the proper motions
motions in the BN/KL outflow.   The rays in the panels  have 
been weighted by the magnitude of the associated velocity vector.  
The image in panel b has been boxcar smoothed by 31 pixels.   The white
cross marks the ejection point of radio sources BN, I, and n.  Each frame is 
150\arcsec\  in extent in the east-west  and  north-south directions.}
\end{figure}
\clearpage

\begin{figure}
\center{  \includegraphics[angle=0,scale=1.0]{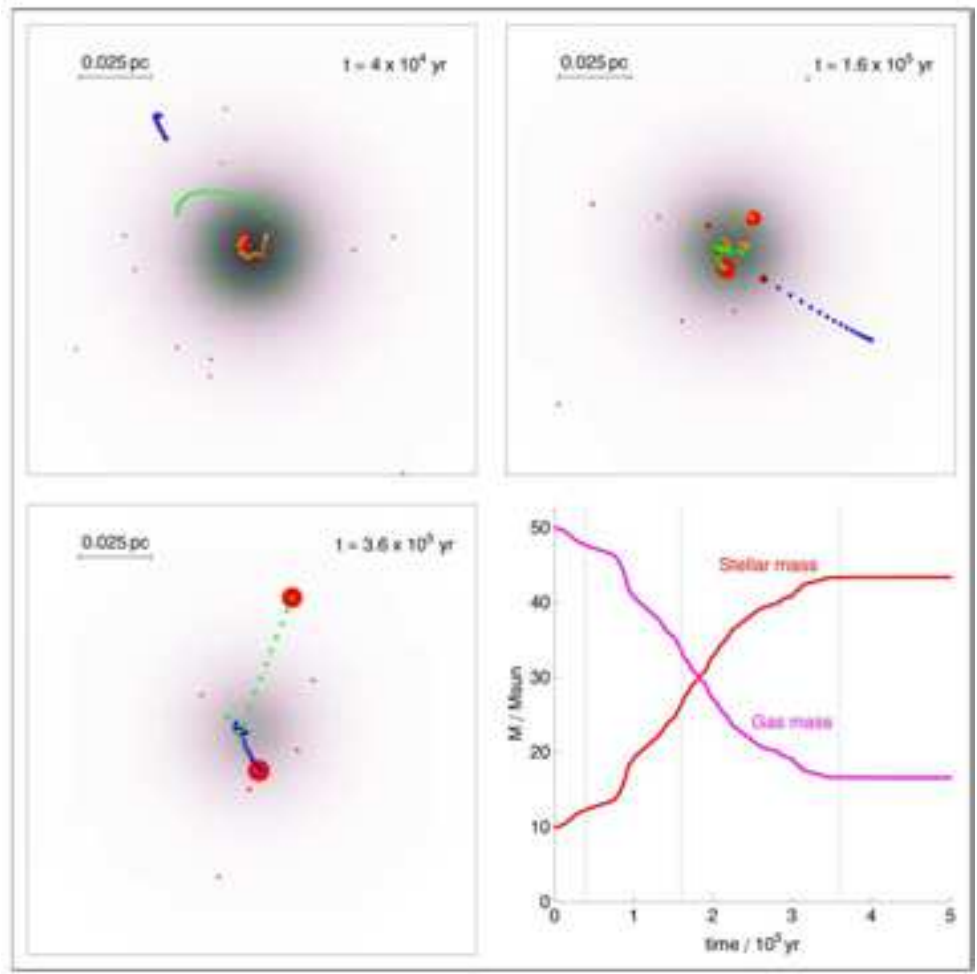}}
\caption{Four frames from a simulation of the formation of
a non-hierarchcal multiple system of massive stars from a spherical clump of
gas.   The simulation starts with 50 M$_{\odot}$ of gas in a Plummer sphere
with a core radius of 0.025 pc.  20 stellar seeds, each having an initial mass of
0.5 M$_{\odot}$ were uniformly distributed in a sphere with radius 0.1 pc, with 
virialized velocities.    Seeds having the lowest speeds and those closest to the
core experience the most rapid growth due to Bondi-Hoyle accretion.  The orbits 
of these stars decay as they gain mass, eventually forming a non-hierarchical 
group near the center of the clump.  Their total mass comes to dominate 
the gravitational potential and the stars are eventually dynamically ejected.  
The fourth panel shows the mass in gas and the mass in
stars, with the times of the snapshots marked with vertical lines.  The 
positions of stars  involved in the final encounter that resulted in
ejection during the last 20 time-steps are shown as colored trails.
}
\end{figure}
\clearpage

\begin{figure}
\center{  \includegraphics[angle=0,scale=1.0]{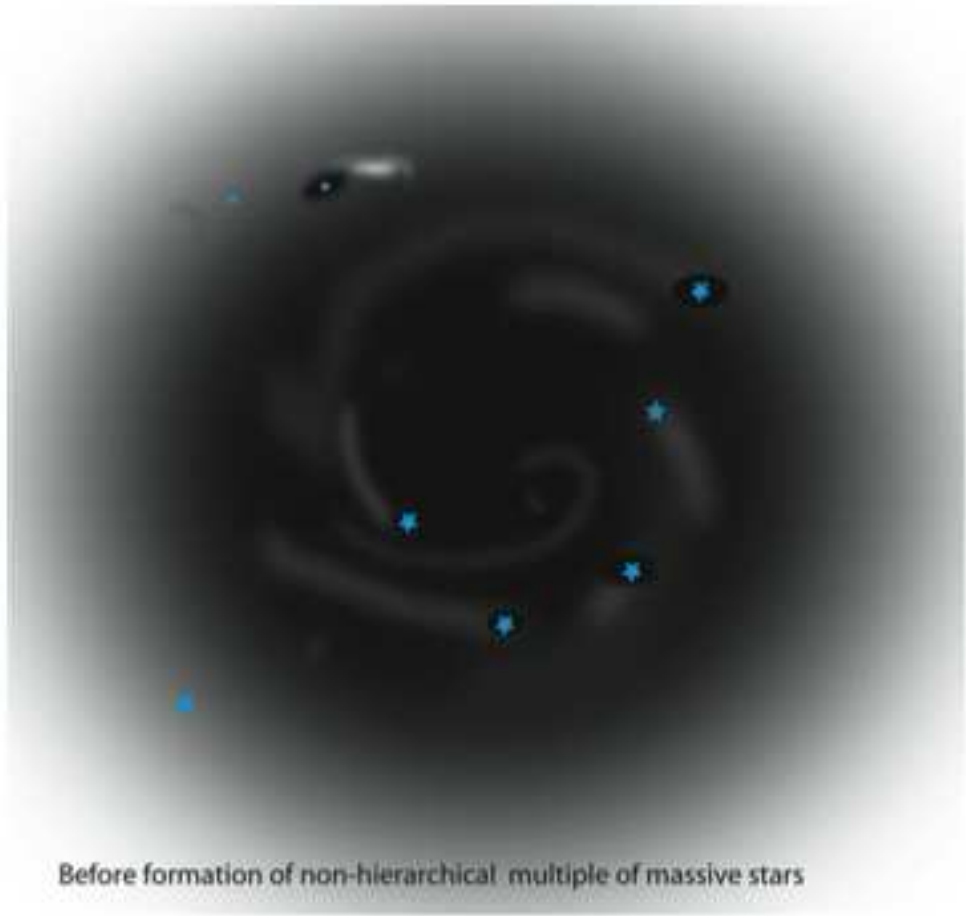}}
\caption{An illustration showing three stages in the evolution of 
OMC1 in the proposed scenario.  
{\bf (a):}  
OMC1 before the formation of a massive non-hierarchical 
multiple system corresponding to the upper-left panel in Figure 4.
{\bf (b):}
OMC1 after the formation of a massive non-hierarchical multiple 
system but before  its decay corresponding to the upper-right
panel in Figure 4.
{\bf (c):}
OMC1 following the ejection of massive stars corresponding to 
the lower-left panel in Figure 4.
}
\end{figure}
\clearpage

\begin{figure}
\center{  \includegraphics[angle=0,scale=1.0]{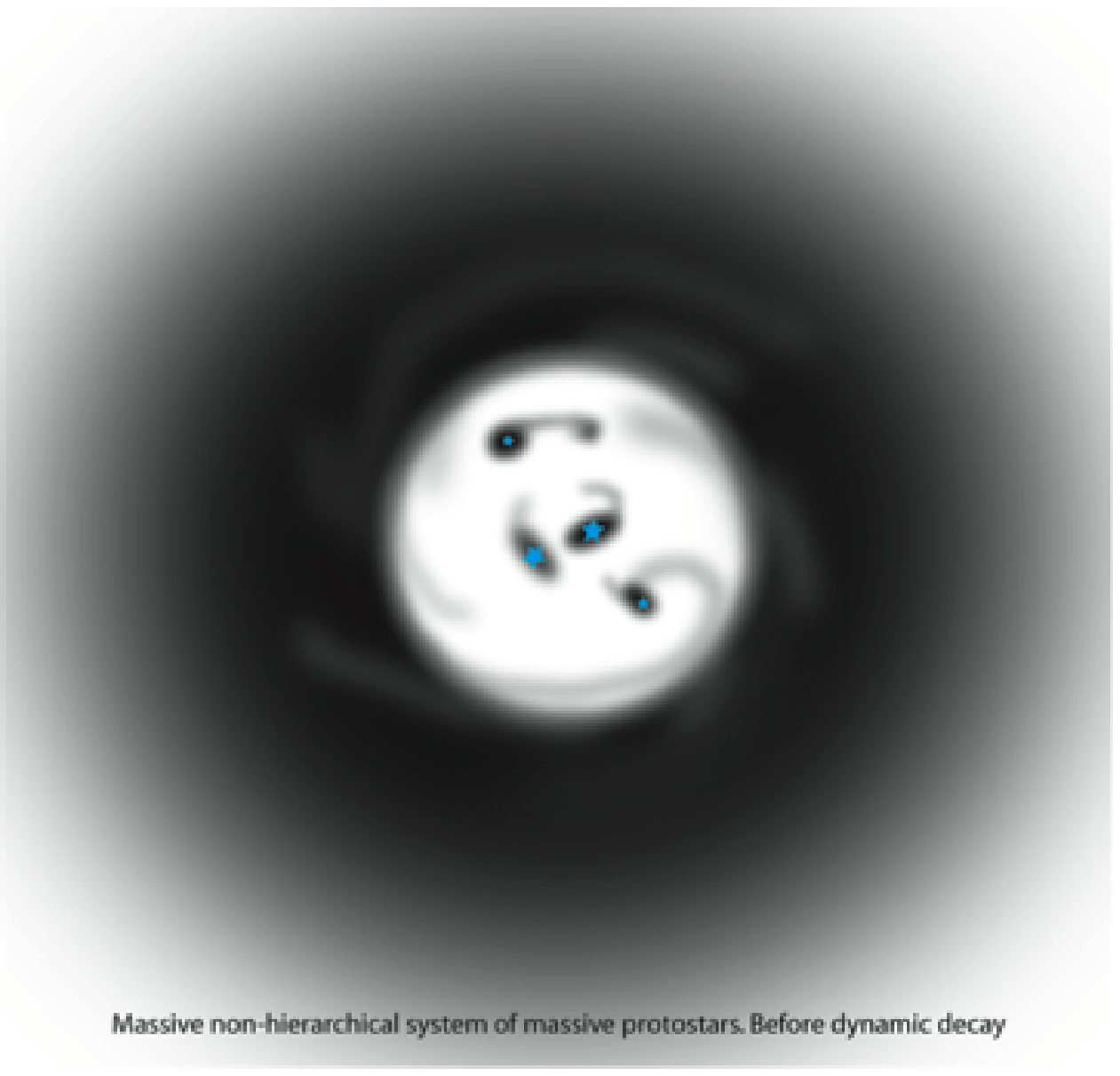}}
\end{figure}
\clearpage

\begin{figure}
\center{  \includegraphics[angle=0,scale=1.0]{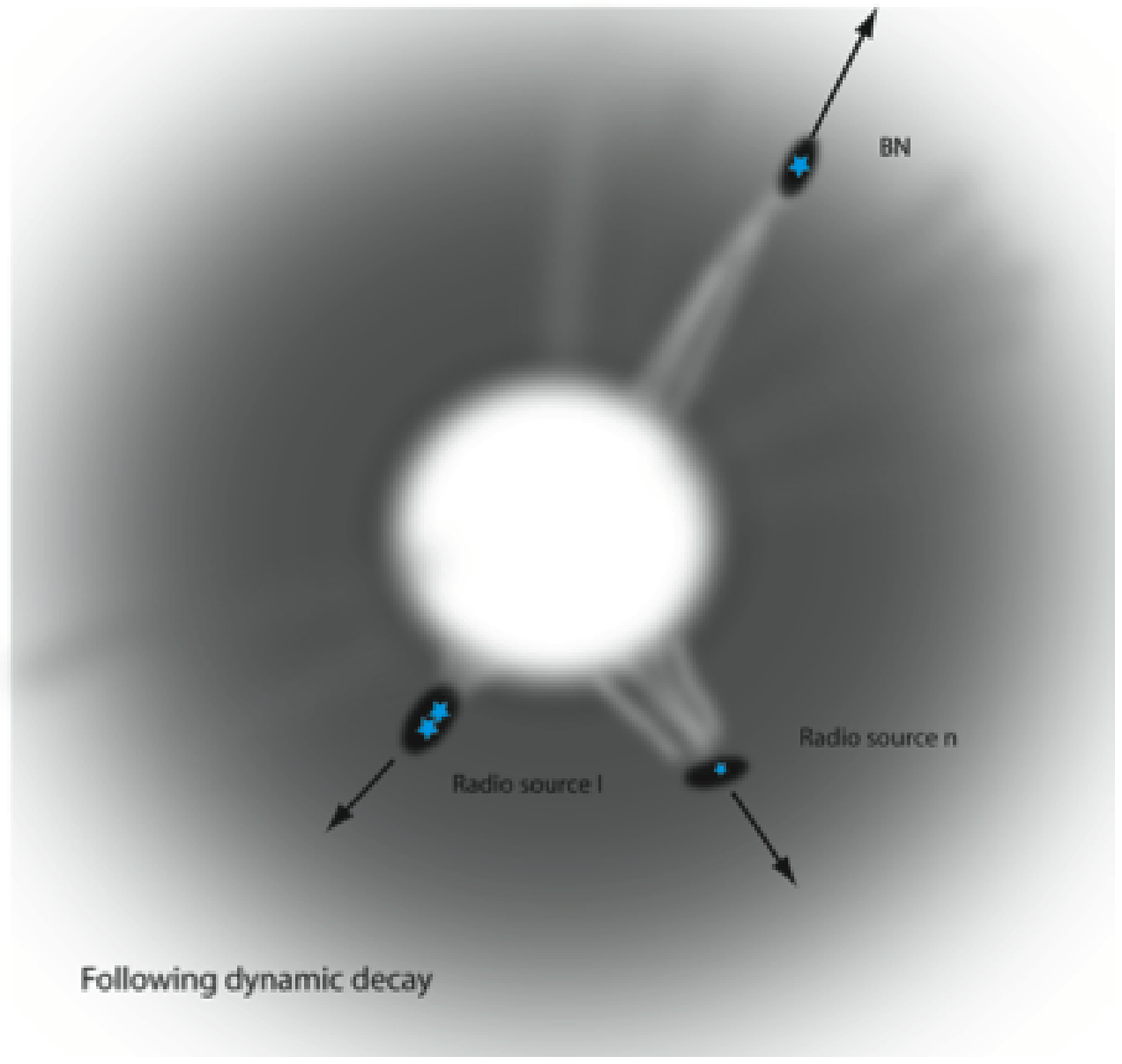}}
\end{figure}
\clearpage

\begin{deluxetable}{llllll}
\tablewidth{0pt}
\tablecaption{  Plausible Kinetic Energies of Ejected Envelopes }
\tablehead{
     \colhead{$\alpha$}           & 
     \colhead{ $R_{out}$}  &
     \colhead{ $R_{in}$ }   &
     \colhead{ $M_{e}$ }   &
     \colhead{ $M_{*}$ }   &
     \colhead{ $E_{excess}$ }  
          \\
     \colhead{}  &
      \colhead{(pc)} &
      \colhead{(AU)} &
      \colhead{(M$_{\odot}$)} &
      \colhead{(M$_{\odot}$)} &
      \colhead{(ergs)} 
      }  
\startdata
\label{table1}
1.5  &     0.1   &   300  &   60   &    40   &   $2.7 \times 10^{45}$   \\
2.0  &     0.1   &   300  &   60   &    40   &   $4.4 \times 10^{45}$   \\
1.5  &     0.1   &   50    &   60   &    40   &   $2.9 \times 10^{45}$   \\
2.0  &     0.1   &   50    &   60   &    40   &   $6.2 \times 10^{45}$   \\
1.5  &     0.1   &   300  &   60   &    20   &   $1.3 \times 10^{45}$   \\
2.0  &     0.1   &   300  &   60   &    20   &   $2.2 \times 10^{45}$   \\
1.5  &     0.1   &   50    &   60   &    20   &   $1.5 \times 10^{45}$   \\
2.0  &     0.1   &   50    &   60   &    20   &   $3.1 \times 10^{45}$   \\
\enddata
\end{deluxetable}

\begin{deluxetable}{llllll}
\tablewidth{0pt}
\tablecaption{  Plausible Kinetic Energies of Ejected Disks }
\tablehead{
     \colhead{$\beta$}           & 
     \colhead{ $R_{out}$}  &
     \colhead{ $R_{in}$ }   &
     \colhead{ $M_{d}$ }   &
     \colhead{ $M_{*}$ }   &
     \colhead{ $E_{excess}$ }  \\
           &
      \colhead{(AU)} &
      \colhead{(AU)} &
      \colhead{(M$_{\odot}$)} &
      \colhead{(M$_{\odot}$)} &
      \colhead{(ergs)}
}  
\startdata
\label{table2}
1.5  &     10   &   0.5   &    0.5   &  10 & $1.9  \times 10^{46}$ \\
1.0  &     10   &   0.5   &    0.5   &  10 & $1.4 \times 10^{46}$ \\
0.5  &     10   &   0.5   &    0.5   &  10 & $1.0  \times 10^{46}$ \\
1.5  &     10   &   0.1   &    0.5   &  10 & $4.4  \times 10^{46}$ \\
1.0  &     10   &   0.1   &    0.5   &  10 & $2.0 \times 10^{46}$ \\
0.5  &     10   &   0.1   &    0.5   &  10 & $1.2  \times 10^{46}$ \\
1.5  &     10   &  0.05  &   0.5   &  10 & $6.2 \times 10^{46}$ \\
0.5  &     10   &  0.05  &   0.5   &  10 & $1.2 \times 10^{46}$ \\
1.5  &     10   &  0.05  &   1.0   &  10 & $1.3 \times 10^{47}$ \\
\enddata
\end{deluxetable}

\clearpage

\end{document}